\documentstyle[12pt,osa]{revtex}
\parskip=\medskipamount

\title{On minimal coupling of the ABC--superparticle \\
to supergravity background}
\author{A.V. ${Galajinsky}^a$\thanks{e-mail address:
galajin@itp.uni-hannover.de; \\
Permanent address:
Department of Mathematical Physics, Tomsk Polytechnical University,
634034, Tomsk, Russia.} and D.M.
${Gitman}^b$\thanks{e-mail address:  gitman@fma.if.usp.br}\\}
\address{$^a$ Institut f\"ur Theoretische Physik, Universit\"at
Hannover,\\
Appelstra{\ss}e 2, D--30167, Hannover, Germany.}

\address{$^b$ Instituto
de F\'\i sica, Universidade de S\~ao Paulo,\\ P.O. Box 66318,
05315-970, S\~ao Paulo, SP, Brasil}
\begin{document}
\maketitle

\begin{abstract}
By rigorous application of the Hamiltonian methods we show that the
$ABC$--formulation of the Siegel superparticle admits consistent
minimal coupling to external supergravity. The consistency check
proves to involve {\it all} the supergravity constraints.

\end{abstract}

\noindent
{\bf PAC codes:} 12.60.J, 04.65.+e

Recently, there has been considerable interest in the study of
the superparticle (superstring) theories due to Siegel [1--7].
The first formulation of such a kind (the $AB$--model) can be viewed
as the conventional superparticle [8] with only first--class
constraints retained [9]. The second modification (the
$ABC$--superparticle) appears if one tries to close the algebra of the
quantum $A,B$--constraint operators in the presence of external super
Yang--Mills [10]. The third reformulation (the $ABCD$--, or
``first--ilk''--superparticle) originated from the attempt to cure problems
intrinsic in the $BRST$ quantized $ABC$--model [11,2]. Having advantage
of being free of problematic second--class constraints, the last two
theories were proven to be physically equivalent to the conventional
superparticle [12], thus suggesting an interesting alternative to attack
the covariant quantization problem intrinsic in the original
superparticle.

An important characteristic feature of the conventional superparticle,
superstring theories is that they admit consistent (i.e. preserving
local symmetries of the free theory) coupling to super Yang-Mills,
supergravity backgrounds [13--15].  In particular, this allowed one to
construct low energy effective action for the superstring theory within
the framework of the sigma--model approach [16] and to get an elegant
geometric interpretation of the super Yang--Mills, supergravity
constraints themselves [14,15]. It is natural then to ask about the
behaviour of the Siegel superparticles in external background
superfields. For the $ABCD$--model in a curved superspace this question
was previously addressed in Ref. [17], where it was proven that the
system {\it can not} be minimally coupled to the supergravity
background, thus showing a serious drawback of the theory.

In this brief note we address the similar question for the
$ABC$--superparticle. As shown below, this model does admit
consistent coupling to external supergravity.
Interesting enough, the consistency check
essentially involves {\it all} the supergravity constraints. This is
in contrast to the conventional superparticle for which a smaller
set is known to be sufficient to define consistent coupling [14,15].

A conventional way to couple a superparticle model to a curved superspace
is to rewrite its action in terms of the vielbein of a flat superspace and
then set the latter to be that of a curved superspace.
This defines the so--called minimal coupling.
For the case at
hand this yields\footnote{For simplicity in this work we examine the
problem in $d=4$ superspace.}
\begin{eqnarray}
&S=\int d\tau \{ \frac 1{2e}({\dot z}^N {e_N}^a
(z)+i\psi\sigma^a\bar\rho-i\rho\sigma^a\bar\psi)^2\cr
&-{\dot z}^N {e_N}^{\alpha} (z)\rho_\alpha
-{\dot z}^N {e_N}_{\dot\alpha} (z){\bar\rho}^{\dot\alpha}
+\rho\sigma^a\bar\rho\Lambda_a\cr
&+\rho^2 k+{\bar\rho}^2 \bar k \}.
\end{eqnarray}
World indices appear on the superspace coordinates $z^M$ and the vielbein
${e_N}^A (z)$ only, all other indices being tangent ones. It is worth
mentioning that minimal coupling may happen to be not sufficient when
examining spinning superparticles [18] in background
superfields. Some other contributions like the magnetic moment coupling
known for the model of spinning particle [19] may turn out to be
necessary.

Because consistent coupling has to preserve a number of
degrees of freedom of the original model\footnote{The conventional
counting of degrees of freedom is known to be problematic for models
involving higher order fermionic constraints. For the case at hand, in
the light--cone gauge there remain two pairs of (complex conjugate)
canonical variables
$(\theta,p_\theta)$, $(\bar\theta,p_{\bar\theta})$.
The quadratic $C$--constraint $p_\theta
p_{\bar\theta}=0$ can consistently be resolved in the original phase
space giving a pair of second class constraints $p_\theta-\alpha
(\theta+\bar\theta)=0$, $p_{\bar\theta}-\bar\alpha
(\theta+\bar\theta)=0$, with $\alpha$ a complex number,
and reproducing the light--cone description of the conventional
superparticle (see also Ref. [12]).}, we pass to the Hamiltonian formalism
and analyze dynamics of the theory. Defining a phase space
momentum to be the left derivative of a Lagrangian with respect to velocity,
one finds the primary constraints
\begin{eqnarray}
&& p_e=0,\cr
&& p_{\psi}=0, \cr
&& p_{\bar\psi}=0, \cr
&& p_{\rho}=0, \cr
&& p_{\bar\rho}=0,\cr
&& p_{\Lambda}=0,\cr
&& p_k=0,\cr
&& p_{\bar k}=0,\cr
&& p_\alpha+\rho_\alpha=0,\cr
&& {\bar p}_{\dot\alpha}+{\bar\rho}_{\dot\alpha}=0,
\end{eqnarray}
where $p_A\equiv {e_A}^N p_N$ and $(p_e,p_{\psi},p_{\bar\psi},
p_{\rho},p_{\bar\rho},p_{\Lambda},p_k,p_{\bar k},p_N)$ are momenta
canonically conjugate to the configuration space variables
$(e,\psi,\bar\psi,\rho,\bar\rho,\Lambda,k,\bar k,z^N)$ respectively.
The canonical Hamiltonian reads
\begin{eqnarray}
&& H^{(1)}=\lambda_e p_e+{\lambda_\psi}^\alpha p_{\psi\alpha}
+\lambda_{\bar\psi\dot\alpha} {p_{\bar\psi}}^{\dot\alpha} \cr
&&+{\lambda_\rho}^\alpha p_{\rho\alpha}+\lambda_{\bar\rho\dot\alpha}
{p_{\bar\rho}}^{\dot\alpha}+{\lambda_\Lambda}^n {p_\Lambda}_n \cr
&&+\lambda_k p_k+\lambda_{\bar k} p_{\bar k}+\lambda^\alpha
(p_\alpha+\rho_\alpha) \cr
&&+{\bar\lambda}_{\dot\alpha}({\bar p}^{\dot\alpha}+
{\bar\rho}^{\dot\alpha})+e\displaystyle\frac{{(p^a)}^2}2
-i\psi\sigma^a\bar\rho p_a \cr
&&+i\rho\sigma^a\bar\psi p_a
-\rho\sigma^a\bar\rho\Lambda_a-\rho^2 k-{\bar\rho}^2 \bar k,
\end{eqnarray}
where $\lambda_{*}$ denote the Lagrange multipliers corresponding to the
primary constraints. In order to analyze consistency conditions for
the primary constraints one introduces the Poisson bracket associated with
the left derivatives [15] (note that under this bracket
$\{ \mu,p_\mu \}=-1$
with $\mu$ a fermion)
\begin{eqnarray}
&\{A,B \}={(-1)}^{\epsilon_A \epsilon_N}
\frac{\overrightarrow{\partial} A}{\partial z^N}
\frac{\overrightarrow{\partial} B}{\partial p_N}\cr
&-{(-1)}^{\epsilon_A\epsilon_B+\epsilon_B\epsilon_N}
\frac{\overrightarrow{\partial} B}{\partial z^N}
\frac{\overrightarrow{\partial} A}{\partial p_N},
\end{eqnarray}
where $\epsilon_A$ is the parity of a function A. In what follows,
the important bracket
\begin{eqnarray}
&\{p_A,p_B \}={T_{AB}}^C p_C-{\omega_{AB}}^C p_C \cr
&+{(-1)}^{\epsilon_A\epsilon_B} {\omega_{BA}}^C p_C,
\end{eqnarray}
proves to be useful.
Here ${T_{AB}}^C$ and ${\omega_{AB}}^C$ are components of the
super torsion and the super connection respectively. The preservation
in time of the primary constraints gives now the secondary ones
(the same as in the flat case)
\begin{eqnarray}
&{(p^a)}^2=0, \cr
&p_a{(\sigma^a\bar\rho)}_\alpha=0, \cr
&{(\rho\sigma^a)}_{\dot\alpha} p_a=0,\cr
&\rho_\alpha {\bar\rho}_{\dot\alpha}=0, \cr
&\rho^2=0, \cr
&{\bar\rho}^2=0,
\end{eqnarray}
and the equations to determine some of the Lagrange
multipliers (together with complex conjugates)
\begin{eqnarray}
&\lambda_\alpha=-ip_a {(\sigma^a\bar\psi)}_\alpha+
\Lambda_a {(\sigma^a\bar\rho)}_\alpha+2k\rho_\alpha, \cr
&\lambda_{\rho\alpha}=({T_{\alpha\beta}}^D p_D-
{\omega_{\beta\alpha}}^\gamma p_\gamma) \lambda^\beta \cr
&-({T_{\alpha\dot\beta}}^D p_D-
{\omega_{\dot\beta\alpha}}^\gamma p_\gamma) {\bar\lambda}^{\dot\beta}\cr
&-e p^a ({T_{\alpha a}}^D p_D+
{\omega_{a \alpha}}^\gamma p_\gamma)\cr
&+i(\psi\sigma^a\bar\rho) {T_{\alpha a}}^c p_c-
i(\rho\sigma^a\bar\psi) {T_{\alpha a}}^c p_c.
\end{eqnarray}
In obtaining Eq. (7) we used the constraints (2),(6) and the explicit
form of the connection $\omega_{Nab}=-\omega_{Nba},
{\omega_{N \alpha}}^\beta=\frac 12 \omega_{Nab}{{(\sigma^{ab})}_\alpha}^\beta,
{\omega_{N \dot\alpha}}^{\dot\beta}=\frac 12 \omega_{Nab}
{{{(\tilde\sigma}^{ab})}^{\dot\beta}}_{\dot\alpha}$. It is worth
mentioning that the last two lines in Eq. (6) follow from the
second and the third ones and, hence, can be omitted. We find it
convenient to keep the corresponding trivial contributions to the
Lagrangian (1) in order to write the local
$\kappa$--symmetry in the simplest form (see Eqs. (12),(13) below).

Consistency conditions for the secondary constraints produce the
equations (together with complex conjugates)
\begin{eqnarray}
&p_a {\sigma^a}_{\alpha\dot\gamma} {\lambda_{\bar\rho}}^{\dot\gamma}
+{(\sigma^a\bar\rho)}_\alpha \{-({T_{a \beta}}^c p_c\cr
&+{\omega_{\beta a}}^b p_b)\lambda^\beta
+({T_{a \dot\alpha}}^c p_c+
{\omega_{\dot\alpha a}}^b p_b){\bar\lambda}^{\dot\alpha}\cr
&+e p^b({T_{ab}}^c p_c+
{\omega_{ba}}^c p_c) \}=0,\cr
&p^a(-{T_{a \alpha}}^D p_D\lambda^\alpha+{T_{a \dot\alpha}}^D
p_D {\bar\lambda}^{\dot\alpha}\cr
&-i(\psi\sigma^b\bar\rho) {T_{ab}}^c p_c+i(\rho\sigma^b\bar\psi)
{T_{ab}}^c p_c)=0,\cr
&\rho_\alpha \lambda_{\bar\rho\dot\alpha}-
{\bar\rho}_{\dot\alpha}\lambda_{\rho\alpha}=0,
\end{eqnarray}
which, after the substitution of Eq. (7), imply further (highly
nonlinear) constraints and, hence, change a number of degrees of
freedom in the problem as compared to that in a flat superspace.
Thus, some restrictions on the background geometry are necessary
to define consistent coupling.
Taking these to be the full set of $d=4, N=1$ supergravity
constraints [20],
\begin{eqnarray}
&{T_{ab}}^c=0,\cr
&{T_{\breve\alpha b}}^c=0,\cr
&{T_{\breve\alpha \breve\beta}}^{\breve\gamma},=0\cr
&{T_{\alpha \beta}}^c=0,\cr
&{T_{\dot\alpha \dot\beta}}^c=0,\cr
&{T_{\alpha \dot\beta}}^c=2i{\sigma^c}_{\alpha \dot\beta},
\end{eqnarray}
where $\breve\alpha$ means either $\alpha$ or $\dot\alpha$, one can
check that equations (8) vanish and, moreover, the
constraints\footnote{The variables
$(e,\psi,\bar\psi,\rho,\bar\rho,\Lambda,k,\bar k)$ together with the
corresponding momenta can be
omitted after imposing the gauge conditions $e=1,\psi=0,\bar\psi=0,
\Lambda=0,k=0,\bar k=0$, and constructing the Dirac bracket
associated with the second class constraints $p_{\rho\alpha}=0,
p_\alpha+\rho_\alpha=0,p_{\bar\rho\dot\alpha}=0,
{\bar p}_{\dot\alpha}+{\bar\rho}_{\dot\alpha}=0$.}
\begin{eqnarray}
&{(p^a)}^2=0, \cr
&p_a{(\sigma^a \bar p)}_\alpha=0, \cr
&{(p \sigma^a)}_{\dot\alpha} p_a=0,\cr
&p_\alpha {\bar p}_{\dot\alpha}=0,
\end{eqnarray}
form a closed algebra and completely determine dynamics of the model
just as in the free case. It is interesting to note that checking this
one essentially
needs to use {\it all} the supergravity constraints (9) as well as the
solutions of the Bianchi identities involving
${T_{a \breve\alpha}}^{\breve\beta}$ (see Ref. [21] for the
explicit relations). This is in contrast to the
conventional superparticle [8] for which the similar analysis shows
that the equations
\begin{eqnarray}
&T_{\breve\alpha (ac)}=\eta_{ac}T_{\breve\alpha},\cr
&{T_{\alpha \beta}}^c=0,\cr
&{T_{\dot\alpha \dot\beta}}^c=0,\cr
&{T_{\alpha \dot\beta}}^c=2i{\sigma^c}_{\alpha \dot\beta},
\end{eqnarray}
with $T_{\breve\alpha}$ an arbitrary superfield, are sufficient to
define consistent coupling (see also Refs. [14,15,22]).

In complete agreement with the Hamiltonian analysis, the Lagrangian (1)
becomes invariant under the local $\kappa$--symmetry when the restrictions
(9) hold. Actually, varying the action (1) with respect to the direct
generalization of the flat $\kappa$--symmetry to a curved superspace
(for technical details see Ref. [22,23])
\begin{eqnarray}
&\delta z^N e_{N \alpha}=-i e^{-1} \Pi_a {(\sigma^a \bar\kappa)}_\alpha,\cr
&\delta z^N e_{N \dot\alpha}=i e^{-1} \Pi_a{(\kappa \sigma^a)}_{\dot\alpha},\cr
&\delta z^N {e_N}^a=i\rho\sigma^a\bar\kappa-i\kappa\sigma^a \bar\rho,\cr
&\delta e=4{\dot z}^N {e_N}^\alpha \kappa_\alpha+
4{\bar\kappa}_{\dot\alpha} {\dot z}^N {e_N}^{\dot\alpha},\cr
&\delta \psi^\alpha=D(\kappa^\alpha),\cr
&\delta {\bar\psi}^{\dot\alpha} =D({\bar\kappa}^{\dot\alpha}),
\end{eqnarray}
where $\Pi^a\equiv {\dot z}^N {e_N}^a+i\psi\sigma^a\bar\rho
-i\rho\sigma^a\bar\psi$ and $D(k^A)$ is the covariant derivative,
and making use of Eq. (9) and the solutions of the Bianchi identities
involving ${T_{a \breve\alpha}}^{\breve\beta}$ [21], one finds that
all the terms entering the variation are proportional to
$\rho\bar\rho,\rho^2,{\bar\rho}^2$, provided the additional variations
of the fields $e,\psi$
\begin{eqnarray}
&\delta e=2e(R \kappa^\alpha \rho_\alpha+\bar R {\bar\rho}_{\dot\alpha}
{\bar\kappa}^{\dot\alpha}\cr
&-\frac 38 \rho^\alpha G_{\alpha\dot\alpha}
{\bar\kappa}^{\dot\alpha}-\frac 38 \kappa^\alpha G_{\alpha\dot\alpha}
{\bar\rho}^{\dot\alpha}),\cr
&\delta \psi^\alpha=-\frac i8 \Pi_a {(\kappa {\sigma}^a)}_{\dot\alpha}
G^{\alpha\dot\alpha}+
\frac {3i}8 \kappa^\alpha \Pi_a {\tilde\sigma}^{a \dot\beta\beta}
G_{\beta\dot\beta},
\end{eqnarray}
have been done. The superfields $R,G_{\alpha\dot\alpha}$ are those
entering the solutions of the Bianchi identities [21].
Obviously, the remnant can be canceled by an appropriate
variation of the fields $\Lambda,k,\bar k$.

Finally, let us briefly comment on the possibility to couple the
$AB$--model to a curved superspace. The Lagrangian to start
with is given by Eq. (1) with the three last terms omitted (the
Hamiltonian analogue is the omitting of the three last lines in Eq. (6)).
Exploiting the same machinery as above, it is easy to check that the
consistency conditions like Eq. (8) {\it do not} vanish even if the full
set of the supergravity constraints holds. They involve terms proportional
to $\rho\bar\rho$ (times background superfields), thus giving further
higher order fermionic constraints in the problem and changing the
original number of degrees of freedom. This suggests that another way
to formulate the $ABC$--superparticle is to require the closure of the
algebra of the $A,B$--constraints in a curved superspace.

To summarize, we conclude that the $ABC$--model is the only one in the
family of the Siegel superparticles which admits consistent minimal
coupling to external supergravity. To our opinion, this is an
indication that the
problem of covariant quantization of the theory, which has
previously been attacked by introducing the $D$--constraint into the
play [11,2], deserves to be re--addressed
without extending the original configuration space.

\vspace{0.3cm}
The work of A.V.G has been supported by DAAD.
D.M.G. thanks CNPq for permanent support.

\end{document}